\documentclass[11pt]{article}
\usepackage{graphicx}
\usepackage{epsfig}
\usepackage{latexsym}
\usepackage{mathrsfs}
\usepackage{feynmf}
\usepackage{balance}
\usepackage{amsfonts}
\usepackage{authblk}

\author[1, 2]{Sharareh Mehrabi Pari \thanks{Email: sharareh.mehrabi.pari@gmail.com}}
\author[1]{Kurosh Javidan \thanks{Email: Javidan@um.ac.ir}}
\author[1]{Fatemeh Taghavi Shahri \thanks{Email: taghavishahri@um.ac.ir}}
\affil[1]{\small{Department of Physics, Ferdowsi University of Mashhad, 91775-1436  Mashhad, Iran} }
\affil[2]{\small{Department of Physics and Technology, University of Bergen, 5007 Bergen, Norway }}
\title{Evolution of heavy quark distribution function in quark-gluon plasmas: using the Iterative Laplace Transform Method }
\date{}

\begin{document}
\maketitle
\begin{abstract}
The "Iterative Laplace Transform Method" is used to solve the Fokker-Planck equation for finding the time evolution of the heavy quarks distribution functions such as charm and bottom in quark gluon plasma. These solutions will lead us to calculation of nuclear suppression factor $R_{AA}$. The results have good agreement with available experiment data from the PHENIX collaboration.      
\end{abstract}

\section*{I. Introduction}

Colliding heavy ions in Relativistic HeavyIion Collider (RHIC) and Large hadron Collider(LHC) experiments generate temperatures many thousand times the temperature of the sun. These collisions are recreating in the laboratory  conditions similar to those just after the big bang. Under these extreme conditions, protons and neutrons melt, freeing the quarks from their bonds with the gluons. This new phase of matter is called quark-gluon plasma. This highly excited state of matter displays properties similar to a nearly perfect fluid, which has been successfully described by hydrodynamic models \cite{i1,i2,i3,i4}.This medium is probed by hard processes with scales ranging from a few $GeV$ to several of $TeV$. The existence of such a phase and its properties are key issues in the theory of quantum chromodynamics (QCD).
Basically, heavy partons, which are produced in early stage of heavy ion collisions, have relatively high density. Such heavy partons could induce a large amount of energy loss for hard patrons produced in the initial stage of the collision. So, the observed jet quenching in such collisions can indicate the formation of this strongly interacting dense matter\cite{i5,i6,i7,i8,i9,i10,i11}. Charm and bottom quarks play crucial role in this endeavor because they are the witness to the entire space-time evolution of the system and due to their large masses, a memory of their interaction history may be preserved\cite{i12,i13,i14,i15}.

Their mass is significantly larger than the temperature of the plasma they are formed in, $M\gg{T_c}$. They are produced in the early stage of heavy ion collisions and they do not dictate the bulk properties of the matter. In addition the heavy quark thermalization time in  perturbative QCD is estimated to be of order of $10-15$ fm/c  and  $25-30$ fm/c  for charm and bottom quarks respectively\cite{i16,i17,i18,i19}. This time is larger than the thermalization time of the light quarks and gluons.

The plasma reaches local equilibrium at an initial temperature $T_i$ and initial thermalization time $\tau_{i}$. It is assumed that its constituent particles are light quarks, anti-quarks and gluons while nonthermalized heavy quarks, which are formed in early stages of collision, are particles with Brownian motion that move through the expanding QGP background. The Fokker-Planck equation (FPE) provides an appropriate framework for studying the evolution of the heavy quark distribution functions \cite{i20,i21}. 

Heavy quarks dissipate their energy during their propagation through QGP via two processes \cite{i22}: 

 The first one is the collisional process, such as ${gQ}\longrightarrow{gQ}$, ${qQ}\longrightarrow{qQ}$,  ${\overline{q}Q}\longrightarrow{\overline{q}Q}$
 
The second one is the radiative process, such as ${Q+q}\longrightarrow{Q+q+g}$, ${Q+g}\longrightarrow{Q+g+g}$ \cite{i23,i24,i25}. 

These processes can be described well by the Fokker-Plank equation.

 In this paper our main task is to study the evolution of heavy quarks in quark gluon plasma by solving the Fokker-Plank equation with new method named "Iterative Laplace Transform Method" \cite{i26,i27}. This method gives numerical solutions in the form of convergent series with easily computable components. To verify our solutions, we calculate the nuclear suppression factor, $R_{AA} (p_T)$, of heavy quarks using simulated data of heavy ion collisions\cite{i28,i29,i30}.

To do this we consider both the collisional and radiation parts of energy loss in FPE and finally we compare our results with experimental data from PHENIX \cite{i31}.

This paper is organized as follows: In the next section we review the Fokker-Planck equation and the collisional energy loss of heavy quarks as well as their energy loss due to radiation. In section III we introduce iterative Laplace transform method. In Section IV we study the time evolution of heavy quarks in QGP using the presented method. Section V is focused on calculating the nuclear suppression factor $R_{AA} (p_T)$. The last section is devoted to concluding remarks.

\section*{II.The heavy quark momentum evolution in FPE dynamics}
The Quark Gluon Plasma (QGP) is formed at the time $\tau_i$ after the impact. Time evolution of the momentum distribution of non-equilibrated heavy quarks due to the interaction with the equilibrated QGP during the time interval ${\tau_i}<\tau<{\tau_{HQ}}$ can be calculated by the Fokker-Planck equation. Evolution of the expanding QGP is described by relativistic hydrodynamics \cite{i32,i33}. 

To derive the Fokker-planck equation, let us start from the Boltzmann transport equation, which in covariant form is \cite{i34}:
\begin{equation}\label{a1}
p^\mu \partial_{\mu}f(x,p)=c\{f\}\quad,
\end{equation}
where $p^\mu$ is the four-momentum of the heavy quarks (HQs), $C\{f\}$ is the collision term and $f(x, p)$ is the momentum distribution function of the heavy quarks.
For uniform plasma, \textit{f} will lose its dependence on \textit{x}, so the Boltzmann equation becomes
\begin{equation}\label{a2}
\frac{\partial f}{\partial t}= \frac{c\{f\}}{E}= (\frac{\partial f}{\partial t})_{coll}
\end{equation}

To simplify the non-linear integro-differential Boltzmann equation, we employ the Landau approximation. This means that we consider soft scatterings with small momentum transfer compared to the particle momentum $p$, and neglect the interaction of heavy quarks with each other. According to these assumptions the evolution of momentum distribution of heavy quarks is given by
\begin{equation}\label{a3}
\frac{\partial f}{\partial t}= \frac{\partial }{\partial p_i}\bigg[ \mathfrak{A}_i(p)f+\frac{\partial }{\partial p_j}[\mathfrak{B}_{ij}(p)f]\bigg]\quad, 
\end{equation}
where'd, $\mathfrak{A}_{i}(p)$ and $\mathfrak{B}_{ij}(p)$ are drag  and diffusion coefficients respectively, defined as:
\begin{equation}\label{a5}
\mathfrak{A}_i=\int d^{3}k \omega(p,k)k_i\quad,
\end{equation}
\begin{equation}\label{a6}
\mathfrak{B}_{ij}= \frac{1}{2}\int d^{3}k \omega(p,k)k_ik_j.
\end{equation}
Here, we considered the elastic scattering of the HQs with gluons, light quarks, and corresponding anti-quarks. All these interactions contribute to determine the collision rate $w (p, k)$ \cite{i35}. 
For an isotropic QGP one can write $\mathfrak{A}_i=p_{i} A(p)$ and $\mathfrak{B}_{i,j}=D(p)\delta_{i,j}$. In the simplest form, we assume that the drag and diffusion coefficients are slowly varying functions of momentum, so Equation (\ref{a3}) reduces to the Fokker-Planck equation:
\begin{equation}\label{a7}
\frac{\partial f(p,t)}{\partial t}=A(p) \frac{\partial}{\partial p}(pf(p,t))+D(p)\bigg[\frac{\partial }{\partial p}\bigg]^2 f(p,t)\quad.
\end{equation}
The drag coefficient is an important quantity carrying information about the dynamics of elastic collisions. It is expected that the drag coefficient should be determined by the properties of the bath and not by the nature of the HQ \cite{i36}.

After heavy ion collision, the charm quarks are produced at a time scale of about $0.07fm/c$. This time is about  $0.02fm/c$ for bottom quarks. These heavy quarks will propagate through the deconfined matter (QGP) while losing energy.

In this work we use the drag coefficient as follows:
 \begin{equation}\label{a9}
A(p,t)=-\frac{1}{p}\frac{dE}{ dL}\quad.
\end{equation}
If coupling between HQs and background QGP is weak, then the diffusion coefficient can be calculated using the Einstein relation, $D(p)=MTA(p)$, where $M$ is the mass of the heavy quark and T is the thermal bath temperature \cite{i37,i38}. The first perturbative estimation of the energy loss in a QGP was proposed by Bjorken for heavy quarks \cite{i39}, which is an important quantity. It may be noted that the jet quenching, drag force, damping rate of particles in the plasma and etc., are determined directly in terms of the energy loss. Furthermore, since the calculated energy loss is proportional to the heavy fermion mass in quark-gluon plasma, it is used as an important tool to identify particles and give more information about the properties of QGP. The collisional energy loss of a heavy quark with energy $E$ and mass $M$ in a thermalized medium with temperature T has been calculated in several papers with different approaches \cite{i40}. In our calculations, the collisional energy loss of HQs in QGP by considering "Hard and Soft Thermal Loops" are given as follows \cite{i41}. 
  
for $E\ll\frac{M^2}{T}$:
\begin{equation}\label{a10}
-\frac{dE}{ dL}= \frac{8\pi\alpha_s^2T^2}{3}\bigg(1+\frac{n_f}{6}\bigg)\bigg [\frac{1}{ v}-\frac{1-v^2}{ 2 v^2}ln(\frac{1+v}{ 1-v})\bigg] ln\bigg[2^\frac{n_f}{ 6+n_f}B(v)\frac{ET}{m_gM}\bigg] \quad,           
\end{equation}
for $E\gg\frac{M^2}{T}$:
\begin{equation}\label{a11}
-\frac{dE}{ dL}= \frac{8\pi\alpha_s^2T^2}{3}\bigg(1+\frac{n_f}{6}\bigg) ln\bigg[2^{\frac{1}{2}(\frac{n_f}{ 6+n_f})}0.92\frac{\sqrt{ET}}{m_g}\bigg]\quad,             
\end{equation}
where $v$ is the HQ speed, $\alpha_s$ is the strong coupling constant, $n_f$ is the number of quark flavors in the medium, $E$ and $M$ are energy and mass of the propagated HQ respectively, $m_g=\sqrt{(1+n_f/6)g^2T^2/3}$ is the thermal gluon mass, $g=\sqrt{4\pi\alpha_s}$ is the gauge coupling parameter and $B(v)$ is a smooth velocity function, which can be taken approximately as $0.7$ \cite{i41}. 

The collisional energy loss of a HQ in a thermalized QGP is calculated by "Hard Thermal Loop" approximation as follows \cite{i42}:
\begin{equation}\label{a12}
-\frac{dE}{ dL}= \frac{8\alpha_s^2T^2}{\pi} \bigg [\bigg(1+\frac{n_f}{3}\bigg) ln\bigg( \frac{ET}{m_{D}^2} \bigg) +\frac{2}{9} ln \frac{ET}{M^2} +0.251 n_{f}+0.747\bigg] \quad,          
\end{equation}
where $m_{D}=\frac{8\alpha_s}{ \pi}(3+n_f)T^2$ is the Debye mass.
Finally the total energy loss due to both collision and radiation precesses is \cite{i43}:
\begin{equation}\label{a13}
-\frac{dE_{total}}{ dL}=\frac{10\alpha_s^2T^2}{3} ln\bigg( \frac{3E}{8\pi\alpha_{s}T}\bigg)\bigg [ ln\bigg( \xi  +\sqrt{1+\xi^2}\bigg)+\xi ln\bigg(\frac{1}{ \xi}  +\sqrt{1+\frac{1}{\xi^2}}\bigg)+\frac{3}{8\pi}\bigg]           
\end{equation}
where $\xi=\frac{9E}{2\pi^{3}T}$.
Using equation $(8)-(11)$, we will calculate the drag and diffusion coefficients for HQ energy loss during the interaction with QGP medium. In the next section we will find the solution of the Fokker-plank equation, eq$(6)$.

\section*{III. The Iterative Laplace Transform Method}

In this section we want to introduce a new method for solving the FP equation, namely the "Iterative Laplace Transform Method" (ILTM) \cite{i44}. Many problems in physics and engineering can be successfully modelled by fractional differential equations (FDE). The FP equation indeed, is an important example of FDEs. Standard methods like finite-difference method are not applicable for such differential equations with non-zero values on the boundaries of initial conditions. Hence should we be looking for an effective way to solve FDEs. This new method was proposed by Daftardar-Gejji and Jafari to seek numerical solutions of nonlinear functional equations. This method is called the Iterative Laplace Transform Method, which is a combination of two powerful methods, namely, the Laplace transform method and the Iterative method. The method gives numerical solutions in the form of convergent series with easily computable components. The most outstanding feature of this method is that it provides an analytical solution using the initial conditions only, without any discretization or restrictive assumptions.

In general, the Fokker-Planck equation can be written as \cite{i44}: 

\begin{equation}\label{b1}
D_t^{\alpha}f=\bigg[D_p^{\beta}A(p,t,f)+D_p^{2\beta}B(p,t,f)\bigg]f(p,t)\quad.
\end{equation}

Here $D_t^\alpha(.), D_p^{\beta}(.), D_p^{2\beta}(.)$ are the Caputo fractional derivative with respect to $t$ and $p$.

In our calculations we use the following definitions:

I. The Caputo fractional derivative of function $(p,t)$ is defined as
 
\begin{equation}\label{b2}
D_t^{\alpha}f(p,t)=\frac{1}{\Gamma(m-\alpha)}\int (t-\eta)^{m-\alpha-1}f^m(p,\eta)d\eta
\end{equation} 

for$m-1<\alpha\leq{m}$,$m\in{N}$.
 
II. The Laplace transform of $f(t)$ is defined as:

\begin{equation}\label{b3}
F(s)=\mathscr{L}[f(t)]=\int_0^{\infty} e^{-st}f(t)dt  \quad.   
\end{equation} 

III. Laplace transform of $D_t^{\alpha}f(p,t)$ is given by

\begin{eqnarray}\label{b4}
\mathscr{L}[D_t^{\alpha}f(p,t)]=x^\alpha\mathscr{L}[f(p,t)]-\displaystyle\sum_{k=0}^{m-1}{f^{(k)}(p,0)x^{\alpha-1-k}}\quad;\quad\,\, m-1<\alpha\leq{m}\quad,
\end{eqnarray}

where $f^{(k)}(p,0)$ is the $k$th derivative of $f(p,t)$ at $t= 0$.

To illustrate how the iterative Laplace transform method works, the general space-time fractional partial differential equation is considered
 
\begin{eqnarray}\label{b5} 
 D_t^{\alpha}f=\mathscr{A}(f,D_p^{\beta}f,D_p^{2\beta}f,...) \qquad & m-1&<\alpha\leq{m}\nonumber
\\&&\!\!\!\!\!\!\!\!\!\!\!\!\!\!\!\!\! n-1<\beta\leq{n}\,\,\,\,\,\,\,  {m,n}\in{N}\quad,
\end{eqnarray} 

where $\mathscr{A}(f,D_p^{\beta}f,D_p^{2\beta}f,...)$ is a linear or nonlinear operator of $f,D_p^{\beta}f,D_p^{2\beta}f,...$. 
    
With the initial condition

\begin{equation}\label{b6} 
f^{(k)}(p,0)=h_k(p);\quad                k=0,1,...,m-1\quad, 
\end{equation}

and $f(p,t)$ will be determined later. 
 
First of all we take the Laplace transform of both sides of $(16)$

\begin{equation}\label{b7} 
 x^\alpha\mathscr{L}[f(p,t)]-\displaystyle\sum_{k=0}^{m-1}{x^{\alpha-1-k}f^{(k)}(p,0)}
=\mathscr{L}[\mathscr{A}(f,D_p^{\beta}f,D_p^{2\beta}f,...)]\quad.
\end{equation} 

After simplification, we have

\begin{equation}\label{b9} 
\mathscr{L}[f(p,t)]=\displaystyle\sum_ {k=0}^{m-1}{x^{-1-k}f^{(k)}(p,0)}+x^{-\alpha}\mathscr{L}[\mathscr{A}(f,D_p^{\beta}f,D_p^{2\beta}f,...)].
\end{equation} 

By inverse Laplace transforms of both sides of Eq. $(19)$ we have:

\begin{equation}\label{b10} 
f(p,t)=\mathscr{L}^{-1}[\displaystyle\sum_{k=0}^{m-1}{x^{-1-k}f^{(k)}(p,0)}]
+ \mathscr{L}^{-1}[x^{-\alpha} \mathcal{L}[\mathscr{A}(f,D_p^{\beta}f,D_p^{2\beta}f,...)]]\quad,
\end{equation} 

and

\begin{equation}\label{b11} 
\mathscr{B}(f,D_p^{\beta}f,D_p^{2\beta}f,...)
= \mathscr{L}^{-1}[x^{-\alpha} \mathscr{L}[\mathscr{A}(f,D_p^{\beta}f,D_p^{2\beta}f,...)]]\quad.
\end{equation} 

The method gives numerical solutions in the form of convergent series 

\begin{eqnarray}\label{b12} 
f(p,t)&=&\displaystyle\sum_{n=0}^{\infty}f_{n},
\end{eqnarray}
and

\begin{eqnarray}\label{b13}\nonumber 
\mathscr{B}
\bigg(\displaystyle\sum_{n=0}^{\infty}f_{n},D_p^{2\beta}\displaystyle\sum_{n=0}^{\infty}f_{n},...\bigg)\nonumber
&&= \mathscr{B}(f_{0},D_p^\beta f_{0},D_p^{2\beta}f_{0})\\\nonumber
&&+\displaystyle\sum_{j=0}^{\infty}\mathscr{B}\bigg(\displaystyle\sum_{k=0}^{j}f_{k},D_p^\beta \displaystyle\sum_{k=0}^{j}f_{k},D_p^{2\beta}\displaystyle\sum_{k=0}^{j}f_{k},...\bigg)\\
&&-\displaystyle\sum_{j=1}^{\infty}\mathscr{B}\bigg(\displaystyle\sum_{k=0}^{j-1}f_{k},D_p^\beta \displaystyle\sum_{k=0}^{j-1}f_{k},D_p^{2\beta}\displaystyle\sum_{k=0}^{j-1}f_{k},...\bigg)\quad.\\\nonumber
\end{eqnarray}
By substituting $(22)$ and $(23)$ into $(20)$ we have

\begin{eqnarray}\label{b14}\nonumber
\displaystyle\sum_{n=0}^{\infty}f_{n}=&\mathscr{L}^{-1}&\bigg[\displaystyle\sum_{k=0}^{m-1}x^{-1-k}f^{(k)}(p,0)\bigg]\\\nonumber
&&+\mathscr{B}(f_{0},D_p^{\beta}f_{0},D_p^{2\beta}f_{0}...)\\\nonumber
&&+\displaystyle\sum_{j=1}^{\infty}\bigg[\mathscr{B}\bigg(\displaystyle\sum_{k=0}^{j}f_{k},D_p^\beta\displaystyle\sum_{k=0}^{j} f_{k},D_p^{2\beta}\displaystyle\sum_{k=0}^{j}f_{k},...\bigg)\\
&&\qquad \qquad \!\! -\mathscr{B}\bigg(\displaystyle\sum_{k=0}^{j-1}f_{k},D_p^\beta \displaystyle\sum_{k=0}^{j-1}f_{k},D_p^{2\beta}\displaystyle\sum_{k=0}^{j-1}f_{k},...\bigg)\bigg]\quad.
\end{eqnarray}

Then we set

\begin{eqnarray}\label{b15}\nonumber 
&&f_0=\mathscr{L}^{-1}\bigg[\displaystyle\sum_{k=0}^{m-1}x^{-1-k}f^{(k)}(p,0)\bigg],\\\nonumber
&&f_1=\mathscr{B}(f_{0},D_p^{\beta} f_{0},D_p^{2\beta}f_{0},...)\\\nonumber
&&.\\\nonumber
&&.\\\nonumber
&&.\\\nonumber
&&f_{m+1}=\mathscr{B}\bigg(\displaystyle\sum_{k=0}^{m}f_{k},D_p^{\beta}\displaystyle\sum_{k=0}^{m}f_{k},D_p^{2\beta}\displaystyle\sum_{k=0}^{m}f_{k},...\bigg)\\
&&\qquad\qquad\!\! -\mathscr{B}\bigg(\displaystyle\sum_{k=0}^{m-1}f_{k},D_p^{\beta}\displaystyle\sum_{k=0}^{m-1}f_{k},D_p^{2\beta}\displaystyle\sum_{k=0}^{m-1}f_{k},...\bigg),m\geq1.\\\nonumber
\end{eqnarray}
Therefore the solution of $(16)$with initial condition $(17)$is given by:

\begin{equation}\label{b16} 
f(p,t)\cong f_0(p,t)+ f_1(p,t)+...+ f_m(p,t)        ,m=1,2,... 
\end{equation}

If we set $\alpha=\beta=1$ in Equation $(26)$ we will reach the FPE. The drag and diffusion coefficients are explicitly timely independent parameters. However we recalculate these parameters before every time steps of our simulation. In this situation we have: 
\begin{equation}\label{b16I} 
f(p,t+\Delta t)\cong f_0+ f_1+...+ f_m  
\end{equation}
where 
\begin{eqnarray}\label{b17I}\nonumber 
&&f_0=f(p,t)\\\nonumber
&&f_1=\Delta t \bigg\{ A(p)\frac{\partial}{\partial p}(pf)+D(p)\bigg[\frac{\partial }{\partial p}\bigg]^2 f \bigg \}_{f=f_0} \quad\\\nonumber
&&.\\\nonumber
&&.\\\nonumber
&&.\\
&&f_{m+1}=\frac{(\Delta t)^{m+1}}{m+1} \bigg\{ A(p)\frac{\partial}{\partial p}(pf)+D(p)\bigg[\frac{\partial }{\partial p}\bigg]^2 f \bigg \}_{f=f_m} ,m\geq1.
\end{eqnarray}
We solved the FPE using both finite difference method and ILTM with different types of initial conditions and compared their results. For initial conditions with non vanishing values at the borderer, the ILTM successfully solves the FPE while results of finite difference method diverges and therefore it failes to find the solution. It may be noted that momentum distribution functions of  HQs, generally find their maximum at $(p=0)$.  

We have taken momentum distribution functions of charm and bottom quarks at $\sqrt{s}=200 GeV$ in Au-Au and also P-P collisions from "The Durham HepData Project" database \cite{i44a} which are simulated data. ILTM has been performed up to $m=2$, which means we calculate $f_0$, $f_1$ and $f_2$ in (\ref{b17I}).

\section*{IV. Time evolution of HQ distribution functions in QGP}

The dissipation coefficients of the thermal bath and initial momentum distribution of HQs, are basic inputs required to solve the FPE. Time evolution of dynamical parameters (like temperature, viscosity and so on) should be taken from suitable models. In our calculation the time dependence of temperature is considered as \cite{i45} 
\begin{equation}\label{b17} 
T(t)=\tau_0^{1/3}T_0/t^{1/3}\quad, 
\end{equation}
where $\tau_0$ and $T_0$ are the initial time and initial temperature that the background of the partonic system has attained in local equilibrium. We have taken $\tau_0=0.33$ fm/c and $T_0=0.373$ GeV in our simulations \cite{i45}. The viscosity effects have not considered for simplification of our problem. Here we considered  a fixed value for coupling constant as $\alpha_s=0.3$. It is clear that the background is not stationary, it expands and cools down with time, so for more accurate results one can consider a strong coupling, $\alpha_{s}$, running with HQ mass and/or the medium temperature. For calculating the time evolution of the HQ momentum distribution by solving the FPE using ILTM, we need three inputs: I) initial distribution function, II) drag coefficient $A(p)$ which can be calculated by Inserting (\ref{a13}) in the Equation (\ref{a9}) and III) the diffusion coefficient as $D=A(p)TM$. Finally, we find the momentum distribution of HQs at transition temperature $T_c=175$ MeV. It is clear that more advanced relations and more suitable assumptions can produce better results which can be found in further works.

Figure 1 demonstrates the drag coefficient for $b$ and $c$ quarks due to collision and radiation separately. This figure clearly shows that the energy loss for $c$ quark is greater than that for $b$ quark. 
\begin{figure}[htp]\label{fig1}
\centerline{\begin{tabular}{cc}
\includegraphics[width=11 cm, height=8 cm]{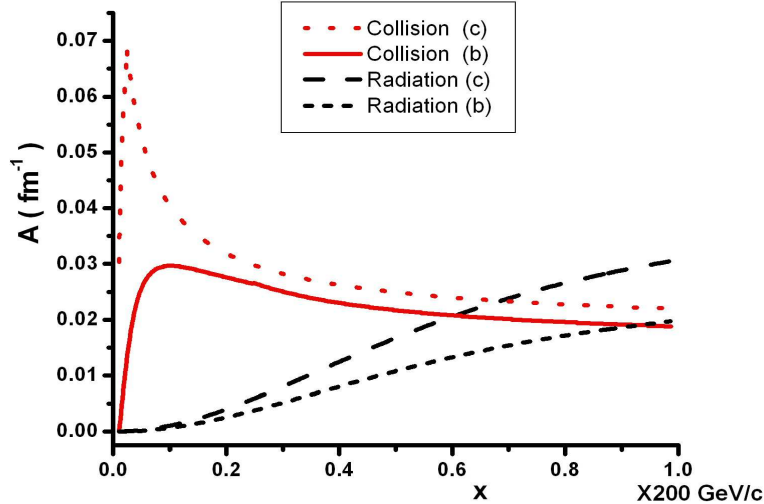}
\end{tabular}}
 \caption{\footnotesize
The drag coefficient due to collision and radiation as functions of momentum for Bottom and Charm quarks in QGP.}
\end{figure}
Figure 2 presents the HQ momentum distributions at initial time ($\tau=0.33$ fm/c) and at final time, when the QGP temperature becomes $T=0.175$ GeV. A general comparison of momentum distributions before and after the evolution indicates that the probability of finding quarks with lower momentums after the evolution is greater than the probability of finding low momentum quarks at initial time. It may be noted that momentum distributions can not be compared directly, because they describe quark distributions at different values of total energy. This figure also shows that energy loss due to suppression of $c$ quark is greater than that for $b$ quark.   
\begin{figure}[htp]\label{fig2}
\centerline{\begin{tabular}{cc}
\includegraphics[width=11 cm, height=8 cm]{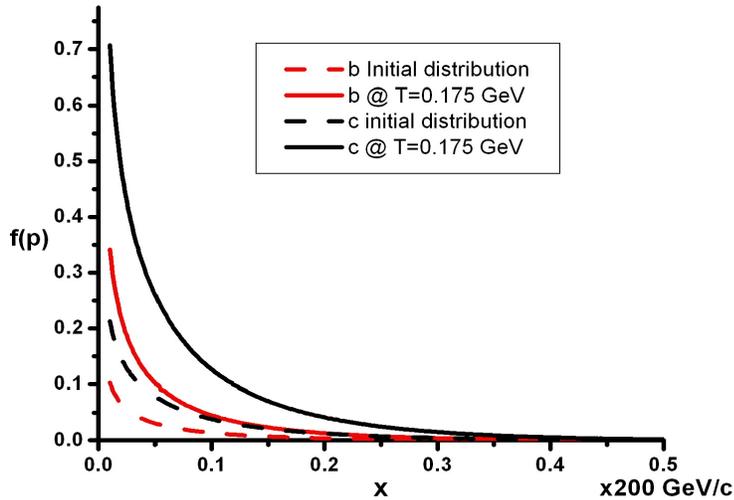}
\end{tabular}}
 \caption{\footnotesize
 Time evolution of b and c quark momentum distributions propagating in QGP medium. Initial values are: $\tau_{0}=0.33 \frac{fm}{c}$ , $T_{0}=0.373 GeV$ and $\alpha_{s}=0.3$. }
\end{figure}

\section*{V. Calculating the nuclear suppression factor $R_{AA}(P_T)$}

One of the most important experimental observables to quantify the nuclear medium is the depletion of high $p_T$ particles ($D$ and $B$ mesons or non-photonic single e$^{-}$ spectra resulting from semi leptonic decays of hadrons containing charm and bottom quarks) produced in Nucleus-Nucleus collisions with respect to those produced in proton-proton collisions which is formulated in nuclear suppression factor $R_{AA}$. The nuclear suppression factor is calculated by solving relativistic hydrodynamic equations for the background medium in QGP phase, simultaneously  with the FP equation for the HQs.The properties of the background bath affect the energy loss of HQs through the drag and diffusion coefficients. The energy loss of HQs is sensitive to initial conditions and the model used to describe the dynamics of the system.

Here the nuclear modification factor $(R_{AA})$ in relativistic heavy ion collisions is calculated for demonstrating the power of ILTM applied to solve the FP equation.

We now proceed with the presentation of our numerical simulation with the formalism described and the inputs mentioned in the previous section. Then we present the comparison between the results obtained with solving FPE by ILTM approach by the experimental data. 
There are several free parameters in the problem, which should be initiated or chosen to find an acceptable fitting the results on the reference data. Equilibration time after the collision, equilibration temperature, contribution of collisional and radiation dissipations on drag and diffusion coefficients, effective light quark flavors $n_{f}$, the charm and the bottom mass, hadron multiplicity at mid-rapidity $\frac{dN}{d\eta}$ and final temperature are some of needed parameters.

To make a realistic connection between simulation results and the experimental data, we have to implement the hadronization mechanisms at the quark-hadron transition temperature in order to find the non-photonic single electron spectra originating from the decays of heavy flavoured mesons (D and B). The solution of the FP equation (described in the previous section) have been convoluted with the fragmentation functions of the heavy quarks to obtain the $P_{T}$ distribution of the heavy mesons. We have used the following heavy quark fragmentation \cite{i45, i46}:
\begin{equation}\label{b18} 
f(z) \propto \frac{1}{z \bigg (z-\frac{1}{z}-\frac{\epsilon}{1-z}\bigg)^{2}}\quad, 
\end{equation}
We have assumed that the electron spectrum produced through semileptonic decay of D and B-mesons is proportional to the heavy meson spectra. Also we have not considered the expansion of the QGP bath. These assumptions create some expected deviation from the experimental data.  

Similar simulations also have been performed to find the electron spectrum created in the P-P collisions. For this purpose HQ distribution functions of proton at $\sqrt{s}=200 GeV$ have been taken as initial conditions to solve FPE again. The ratio of these two quantities is proportional to the nuclear suppression factor, $R_{AA}$ measured in experiments: 
\begin{equation}\label{b19} 
R_{AA}(P_{T})\propto \frac{ \frac{ dN^e }{dP_{T} d\eta}^{Au-Au}} { \frac{ dN^e }{dP_{T} d\eta}^{P-P}} 
\end{equation}
A small value of $R_{AA}$ indicates a strong suppression and therefore, large energy loss of heavy quarks in the medium. It is clear that this ratio is equal to one in the absence of any medium. 
 
Figure 3 presents contribution of collision and radiation energy loss in suppression factor $R_{AA}$ separately. Displayed experimental data has been obtained from the PHENIX collaborations for $Au + Au$ collisions at $\sqrt{s_{NN}} = 200 GeV$.   
\begin{figure}[htp]\label{fig3}
\centerline{\begin{tabular}{cc}
\includegraphics[width=11 cm, height=8 cm]{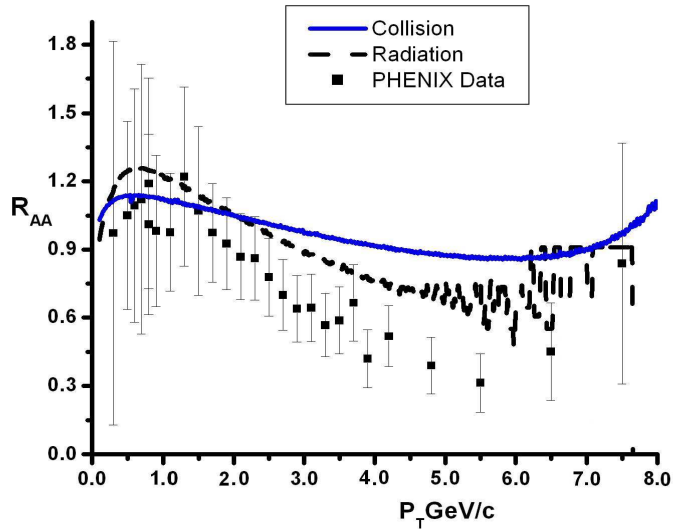}
\end{tabular}}
 \caption{\footnotesize
 Nuclear suppression factor $R_{AA}$ due to collisional and radiation energy loss separately as functions of $P_T$. Used parameters are: $ \tau_{0}=0.33 fm/c, T_{0}=0.375 GeV, \alpha_{s}=0.3, T_{c}=0.175 GeV$.}
\end{figure}

This figure shows that suppression at higher values of $P_{T}$  can not be explained by collisional energy loss. In other words, contribution of radiation in energy loss at higher $P_T$ is more important than collisional energy loss only. Figure 4 demonstrates our best simulation result by considering both collision and radiation energy loss in calculating the drag and diffusion coefficient to solve the FPE. The drag coefficient has been considered as $A_{total}=A_{collison}+1.7A_{radiation}$. This figure shows a very good agreement between results and experimental data at least for $P_{T}<5$ GeV. It may be noted that the final temperature has been taken $T_{c}=0.16$ GeV and an offset 0.05 has been added too. As mentioned before we have not considered some details of the real situation in our simulations.    
\begin{figure}[htp]\label{fig4}
\centerline{\begin{tabular}{cc}
\includegraphics[width=11 cm, height=8 cm]{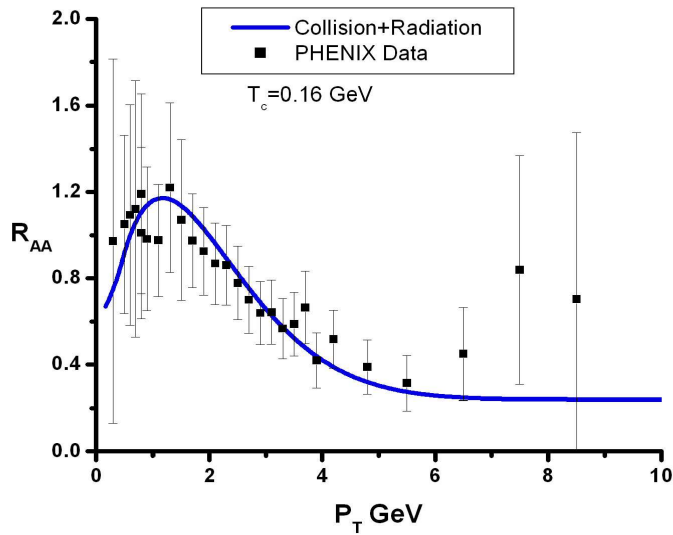}
\end{tabular}}
 \caption{\footnotesize
 Nuclear suppression factor $R_{AA}$ as function of $P_T$. Energy loss due to collision and radiation have been considered in calculating the drag coefficient. }
\end{figure}

\section*{IV. Conclusions and remarks }
The "Iterative Laplace Transform Method" has been introduced as an effective method to solve the Fokker-Planck equation for initial conditions with non-zero values at the boundaries. A numerical algorithm for solving FPE is presented in this paper. Our calculations show that this method is able to solve time evolution of distribution function of HQs successfully. To verify the ability of ILTM and demonstrating its application, we calculated the nuclear suppression factor $R_{AA}$.  It is shown that there is a good agreement between simulation results and reported experimental data. 

 Therefore many problems can be investigated using the ILTM. Effects of a running coupling constant, HQ mass, dead cone effect and LPM effects, QGP viscosity, different equations of state as models of QGP are subjects which can be studied in further works.  

\section*{Acknowledgements}
The authors are very grateful to professor Laszlo Pal Csernai, for many helpful suggestions and comments. S. M. P. also gratefully acknowledges warm hospitality of Institute for Physics and Technology, University of Bergen during her visit. This work is supported by the Ferdowsi university of Mashhad with grant NO. 3/27969-1392/10/15.

\end{document}